# Monte Carlo simulation of molecular dynamics for amino acid production from carboxylic acid via $^{14}$C β-decay

Tomonori Fukuchi[1]*, Takashi Niwa[2], Takamitsu Hosoya[2,3], Yasuyoshi Watanabe[1]

[1] Laboratory for Pathophysiological and Health Science, RIKEN Center for Biosystems Dynamics Research, Kobe, Hyogo, Japan

[2] Laboratory for Chemical Biology, RIKEN Center for Biosystems Dynamics Research, Kobe, Hyogo, Japan

[3] Laboratory of Chemical Bioscience, Institute of Biomaterials and Bioengineering, Tokyo Medical and Dental University (TMDU), Chiyoda-ku, Tokyo, Japan

*Corresponding author
E-mail: tfukuchi@riken.jp

## Abstract

To study the chemical origin of life, we investigated the amino acid production probability from radiocarbon ($^{14}$C)-containing carboxylic acid via β-decay using a newly developed Monte Carlo simulator for molecular dynamics calculation. Using the simulation, we calculated the dynamical trajectory of $^{14}$N recoiled via $^{14}$C β-decay in [3-$^{14}$C]propionic acid and evaluated the staying probability of $^{14}$N in the compound, which is the glycinium (a protonated glycine) production probability from [3-$^{14}$C]propionic acid. The glycinium production probability was calculated to be 78.7% assuming static compound structures of [3-$^{14}$C]propionic acid and a glycinium C–N binding potential energy of 3.93 eV. From calculations with various binding potential parameters, a glycinium production probability of approximately 30% was expected by simulation in case of a loose C–N binding potential energy of 2-eV.

## Introduction

The abiotic synthesis of amino acids in the first step of the chemical origin of life remains a fundamental, unsolved subject. Oparin gave the first theoretical answers regarding this subject [1] and Urey and Miller gave experimental results [2–4]. These answers involve a synthesis of amino acids via an electric discharge under a hypothetical reducing atmosphere of the primitive Earth: methane (CH$_4$), ammonia (NH$_2$), water (H$_2$O), and hydrogen (H$_2$). Although a recent study revealed that the primitive Earth consisted of a weakly reducing atmosphere, including nitrogen ($N_2$), carbon dioxide ($CO_2$), carbon monoxide (CO), and water ($H_2O$), this composition differed from the hypothetical atmosphere used for the Urey and Miller experiments [5]. Nevertheless, the electric discharge synthesis of amino acids confirmed under such an atmosphere is a highly promising hypothesis [6]. Moreover, the discovery of amino acids on meteorites supports the hypothesis of extraterrestrial origin of amino acids [7-10]. Since homochirality was confirmed in the amino acid of the Murchison meteorite in 1982 [11], this amino acid's origin was discussed with bio-homochirality.

Recently, another hypothesis was proposed that the intense impacts of extraterrestrial objects produce amino acids, supported by experimental results with a high-velocity impactor [12, 13].

In this paper, we propose another possibility of the origin of amino acid: conversion of radiocarbon ($^{14}C$) in the methyl group of a carboxylic acid into amino acid, i.e., the [$^{14}C$]methyl group is converted into an amino group with $^{14}C$ β-minus-decay, producing $^{14}N$ with a half-life of 5700±30 years [14]. In $^{14}C$ β-decay, daughter nuclide $^{14}N$ has momentum by the recoil of the emitted β-electron and antineutrino. If daughter nuclide $^{14}N$ binds with a compound despite the recoil momentum, an amino acid is produced from the $^{14}C$-containing carboxylic acid.

On the present Earth, $^{14}C$ is mainly produced from $^{14}N$ by cosmic rays with the $^{14}N(n, p)^{14}C$ reaction [15], and the natural abundance ratio on carbon is about one part per trillion, which varies by solar activity and nuclear-weapon testing done recently [16].

Since the natural abundance of $^{14}C$ or natural amount of carboxylic acids in the primitive Earth is unknown, even if an amino acid can be produced from a $^{14}C$-carboxylic acid, whether those amounts are sufficient to be the origin of life (or a part of it) is unknown and necessary to discuss. In addition, the long half-life of $^{14}C$ must be a part of such a discussion. The unnecessity of nitrogen for the synthesis of carboxylic acid may be an advantage for the origin of amino acid; experimental results from the high-velocity impactor [12, 13] show higher productivity of carboxylic acid than amino acid. However, in this study, we limit our discussion to the production probability of an amino acid from a $^{14}C$-carboxylic acid by $^{14}C$ β-decay.

As the simplest example, we studied the production probability for glycinium—a protonated form of glycine—which is the staying probability of recoiled $^{14}N$ in the $^{14}C$ β-decay of the compound by orbital analysis of recoiled $^{14}N$ using Monte Carlo initial momentum determinations and molecular dynamics calculations.

## Material and Methods

To evaluate the staying probability of recoiled $^{14}N$ in a $^{14}C$ β-decay, we developed a Monte Carlo simulation code to calculate the recoiled molecular dynamics following Newton mechanics. In developed code, the production probability of glycinium is simulated from [3-$^{14}C$]propionic acid as the simplest example. In this paper, [3-$^{14}C$]propionic acid denotes a propionic acid with $^{14}C$ in the terminal methyl group.

Figure 1 shows schematic illustrations of glycinium production and nonproduction in the β-decay of [3-$^{14}$C]propionic acid. For the recoiled $^{14}$N that stayed in the compound, glycinium was produced from [3-$^{14}$C]propionic acid (Fig. 1, top). Meanwhile, for the recoiled $^{14}$N that escaped from the compound via homolysis of the $^{14}$C–N bond, [3-$^{14}$C]propionic acid resolved into a carboxymethyl radical and an aminyl radical (Fig. 1, right bottom).

In β–minus–decay with maximum β–ray energy $E_0$, the β-ray energy $(E)$ distribution is represented with emitting electron momentum $p$ and total β-electron energy $W$ using the following equation:

$$N(E) = pW(E_0 - E)^2 F(Z,W)C(E),$$

where $Z$, $F(Z, W)$ and $C(E)$ are the atomic numbers of the parent nuclide, Fermi function, and spectrum correction factor, respectively. Electron momentum $p$ and total energy $W$ are expressed, respectively, as follows:

$$p = (W^2 - 1)^{\frac{1}{2}}$$

$$W = \frac{E + m_e c^2}{m_e c^2},$$

where $m_e c^2$ is the resting mass of the electron. β-decay of $^{14}$C is a Gamow–Teller transition, and spectrum correction factor $C(E)$ in this type of transition equals 1. The Fermi function is calculated by the following equations using the Gamma function, as follows:

$$F(Z,W) = 2(1+\gamma)(2pR)^{2\gamma-2} e^{\pi y} \frac{|\Gamma(\gamma + iy)|^2}{(\Gamma(2\gamma + 1))^2}$$

$$\gamma = (1 - (\alpha Z)^2)^{\frac{1}{2}}$$

$$y = \alpha \frac{ZW}{p},$$

where $\alpha$ is a fine structural constant and $R$ is a nuclear radius ($R = 1.43 \times 10^{-13} A^{1/3}$ m, and A is a nuclear mass in atomic unit) [17-19].

The Gamow–Teller β-decay nuclide, including $^{14}$C, has axial-vector type β-neutrino (antineutrino) angular correlations expressed as follows:

$$W(\theta) = 1 - \frac{1}{3}\frac{v}{c}\cos\theta,$$

where $v$ and $\theta$ are the neutrino (antineutrino) velocity and the polar angle of the emission direction with respect to the β-ray emission direction [20, 21].

Using these equations for energy distributions $N(E)$ and the angular correlations of β-electron and antineutrino $W(\theta)$, the initial recoil energy and direction of daughter nuclide $^{14}$N in the $^{14}$C β-decay are determined event-by-event using the randomly generated β-electron and the antineutrino emission energies and directions. The emitting β-ray energy and NH$_3$ recoil energy distributions with 1,000,000

Monte Carlo trials are shown in Fig. 2.

Molecular dynamics calculation for determining initial recoil momentum and direction of $^{14}$N was performed to evaluate C-N binding probability despite recoil momentum. Moreover, because the energy of N–H binding in the amino group is nearly equivalent to that of C–N binding, there is a possibility of cleavage of N–H binding instead of the cleavage of C–N binding. However, the maximum $^{14}$N recoil energy of about 5 eV results limiting a single cleavage of N–H binding, and in that case, N remains in the compound to form an amino group, that is the production of glycine structure. Accordingly, the recoiled $^{14}$N is considered to escape as a radical cation form of an aminyl cation radical ($^{+\cdot}$NH$_3$) or that of a carboxymethylaminyl radical (R-$^{+\cdot}$NH$_2$), and we assumed that recoiled $^{14}$N is accompanied by three hydrogen atoms (NH$_3$) in our simulation.

The flight orbit of the recoiled NH$_3$ is calculated using a four-order Runge-Kutta stepper with 1 attosecond step according to Newton mechanics. The simulation assumes a Morse-type potential for the C–N covalent binding and calculates the probability of NH$_3$ remaining or escaping from the C–N binding potential by recoiled NH$_3$ orbit calculation. The Morse potential is expressed as follows:

$$V(r) = D_e\left(e^{-2a(r-r_0)} - 2e^{-a(r-r_0)}\right),$$

where $D_e$ and $r_0$ are the potential depth and the equilibrium bond distance (distance from the potential center to the potential minimum). Parameter $a$ is defined as

$$a = \sqrt{\frac{k_e}{2D_e}}$$

by spring constant $k_e$, which defines the potential width. The first term of the Morse potential represents the repulsive force in a short range, and this force is applied to all atoms of the glycinium.

The molecular configurations of propionic acid and glycinium were calculated via a density functional theory (DFT) method in the Gaussian 16 package [22]; and these molecular coordinates are listed in Table 1. From these configurations, the distances between C–$^{14}$C in propionic acid and C–N bond in glycinium were 153.2 and 150.3 pm, respectively. The 2.9-pm difference in distance in the potential minimum results in a difference between the initial position of the recoiled NH$_3$ and potential minimum because the recoiled NH$_3$ started from outside the potential minimum in the glycinium.

Morse potential parameters $D_e$, $r_0$, and $k_e$ for C–N bond in glycinium were also determined via DFT calculations to be 3.93 eV, 150.2 pm, and 0.00268, respectively. We performed orbital simulations by these parameters while varying them.

**Table 1.** Molecular structure of propionic acid (a) and glycinium (b) calculated using Gaussian code. Third position of carbon was set to origin and the positions of $^{14}$C in propionic acid and $^{14}$N in glycinium are aligned in X-axes.

(a) Propionic acd

|   |     | X (pm) | Y (pm) | Z (pm) |
|---|-----|--------|--------|--------|
| 1 | $^{14}$C | 153.2 | 0 | 0 |
| 2 | H | 192.4 | 97.7 | 28.7 |
| 3 | H | 192.3 | -25.2 | -98.9 |
| 4 | H | 190.5 | -74.2 | 70.9 |
| 5 | C | 0 | 0 | 0 |
| 6 | H | -39.4 | -98.1 | -26.6 |
| 7 | H | -37.0 | -26.0 | 99.6 |
| 8 | C | -57.8 | 98.8 | -97.9 |
| 9 | O | -126.4 | 72.1 | -192.9 |
| 10 | O | -23.4 | 225.7 | -67.8 |
| 11 | H | -62.9 | 282.5 | -134.9 |

(b) Glycinium

|   |   | X (pm) | Y (pm) | Z (pm) |
|---|---|--------|--------|--------|
| 1 | N | 150.3 | 0 | 0 |
| 2 | H | 187.0 | 52.1 | 80.2 |
| 3 | H | 187.2 | 46.2 | -83.6 |
| 4 | H | 187.7 | -95.1 | 3.3 |
| 5 | C | 0 | 0 | 0 |
| 6 | H | -34.6 | -54.6 | -87.7 |
| 7 | H | -34.4 | -51.6 | 89.6 |
| 8 | C | -57.4 | 141.1 | -2.1 |
| 9 | O | -174.1 | 162.0 | -4.9 |
| 10 | O | 40.9 | 232.3 | 0.0 |
| 11 | 3.5 | 321.6 | 321.6 | -1.0 |

## Results

Before evaluating the production ratio of glycinium from [3-$^{14}$C]propionic acid, to make a case for the binding or escaping events of recoiled NH$_3$ in the compound, 2-dimensional plots for the direct distances of recoiled NH$_3$ versus curvature ratios are illustrated at three elapsed times: 50, 100, and 150 fs (Fig. 3). The curvature ratio is defined as the ratio between the direct and orbital distances of the recoiled NH$_3$. In these plots, the calculated events are divided into two groups at a time development of 150 fs. One group of events, which has a small curvature stayed around the origin. The other was for escaping far away with large curving ratios. To define these groups, we set a threshold of 0.7 on the curving ratio (Fig. 3 (c)).

Figure 4 shows the 2-dimensional projection of the NH$_3$ flight orbits for the first 100 events of the group below and above threshold 0.7 in the calculation of the 150-fs duration. The plots clarify that the former group (Fig. 4 (a)) consists of events that stayed around the original position, and these were considered to bind to the compound and formed glycinium. Moreover, the latter group (Fig. 4 (b)) consists of events that escaped from the original positions, and in these events, NH$_3$ escaped from the compound. We named these groups "binding" and "escaping," respectively. The binding and escaping ratios were evaluated using the events below and above the 0.7 threshold in NH$_3$ orbital calculations for the 150-fs duration.

We performed 10,000 Monte Carlo trials for each parameter condition. The glycinium production probability with the Morse potential parameters for the static structure ($D_e$ = 3.93 eV, $r_e$ = 150.3 pm, $k_e$ = 0.00268) was calculated to be 78.7%. The glycinium production probabilities with changing the Mores potential parameters were also calculated. The calculated results with the variation of the Morse potential parameters are shown in Fig. 5. Figure 5 (a) shows the potential depth ($D_e$) dependences of the glycinium production probability. The glycinium production probability with basic Morse parameter is represented by red circles in Fig. 5. Because the depth of the binding potential, the glycinium's static structure of propionic acid calculated via DFT calculation (Table 1), gives an upper limit of the potential depth, the calculations were performed by varying the potential depth in the shallower side.

The production ratios with variations of parameters $r_e$ and $k_e$ are shown in Figs. 5 (b) and (c). In these calculations, we used the parameter values with a static structure of glycinium, except for the varying parameters.

Figure 6 shows the glycinium production ratios with a function of the NH$_3$ recoil energies. The calculations were performed for three potential depths: $D_e$ = 3.9, 3.5, and 3.0 eV. Parameters $r_e$ and $k_e$ were set to the values for the static structure of glycinium ($r_e$ = 150.3 pm, $k_e$ = 0.00268).

## Discussion

From the simulation results with the binding potential parameters of the glycinium calculated via a DFT method, the largest part (78.7%) of the recoiled NH$_3$ stayed in the compound, which was the high production ratio of glycinium from the [3-$^{14}$C]propionic acid. This ratio was almost unchanged by the parameter variations in the equilibrium C–N bond distance ($r_e$) and potential width ($k_e$) (Figs. 5 (b)). However, the production ratio was strongly dependent on the potential depth ($D_e$) (Fig. 5 (a)).

From the glycinium production probability as a function of the NH$_3$ recoil energy in Fig. 6, only the glycinium production probability for the recoiled NH$_3$ with the energy of the upper and lower 0.5 eV of the potential depth depended on the initial direction of the recoil momentum. Because the distribution of the NH$_3$ recoil energy was physically unchanged, the glycinium production probability strongly depended on the binding potential depth. The β-ray energy distribution in which lower energy events appear with high frequency caused a rapid decrease in the glycinium production probability in shallow depth of C–N binding potential.

We used the parameters for the static molecular structure calculated via a DFT method in Gaussian 16 package for our simulations. However, since the structures of the actual molecules were dynamically changed, the potential depth became shallower than that of the static molecules owing to the shape change of the compound, thermal motion, and changes of the orbital electron configuration. The position difference between the original position of the recoiled $^{14}$N in propionic acid and the potential minimum of the C–N binding in the glycinium caused the starting of recoiled $NH_3$ from nonminimum position of potential. Therefore, the calculated production rate of 78.7% is the upper limit for glycinium production. However, assuming a typical C–N covalent binding energy around 3 eV [23], approximately 57% of the event stayed and produced glycinium (Fig. 5 (a)). From Fig. 5 (a), in the case of loose binding for example with a 2-eV potential depth, more than 30% of the [3-$^{14}$C]propionic acid was converted into glycinium via $^{14}$C β-decay.

## Conclusions

Our calculations provide an approximation of glycinium production probability from [3-$^{14}$C]propionic acid by $^{14}$C β-decay. For further studies, simulation with a dynamical molecular structure is required. Experimental quantification is also required to confirm the amino acid production from $^{14}$C-carboxylic acids. We will experimentally confirm the production of glycinium from [3-$^{14}$C]propionic acid. Assuming a 50% production ratio of glycinium, 1.6 nmol of glycinium could be produced in 60 days from 37 MBq (15.9 mmol) of [3-$^{14}$C]propionic acid. This amount of produced glycinium from the highly purified [3-$^{14}$C]-propionic acid can be measured by a current analysis method, such as LC-MS. Our future work includes a verification experiment.

If glycinium is produced from [3-$^{14}$C]propionic acid via $^{14}$C β-decay, amino acids other than glycinium can also be produced from various $^{14}$C-incorporated carboxylic acids. Assuming all or some amino acids in life originate from $^{14}$C β-decay in carboxylic acid, the parity conservation that breaks in $^{14}$C β-decay might be related to the amino acid homochirality in life.

## Acknowledgments

The authors wish to thank members of RIKEN Center for Biosystems Dynamics Research for helpful discussions. This work was partially supported by JSPS KAKENHI Grant Number JP20K12704.

## Author Contributions

Conceptualization: Tomonori Fukuchi.
Software: Tomonori Fukuchi.

Analyzed data: Tomonori Fukuchi.

Gaussian calculation: Takashi Niwa.

Supervision: Takamitsu Hosoya, Yasuyoshi Watanabe.

Writing – original draft: Tomonori Fukuchi

Writing – review & editing: Takashi Niwa, Takamitsu Hosoya, Yasuyoshi Watanabe

## References

1. Oparin AI. The Origin of Life. Macmillan Co., New York 1938.

2. Urey HC. On the Early Chemical History of the Earth and the Origin of Life. Geophysics 1952; 38: p.351-363.

3. Miller SL. A Production of Amino Acids Under Possible Primitive Earth Conditions. Science. 1953; 117: p. 528-529.

4. Miller SL, Urey HC. Organic Compound Synthes on the Primitive Earth: Several questions about the origin of life have been answered, but much remains to be studied. Science. 1959; 130: p. 245-251.

5. Sossi PA, Burnham AD, Badro J, Lanzirotti A, Newville M. O'Neill J St.C. Redox state of Earth's magma ocean and its Venus-like early atmosphere. Sci. Adv. 2020; 6: eabd1387.

6. J. L. Bada. New insights into prebiotic chemistry from Stanley Miller's spark discharge experiments. Chem. Soc. Rev. 2013; 42: p.2186-2196.

7. Kvenvolden KA et al. Evidence for Extraterrestrial Amino-acids and Hydrocarbons in the Murchison Meteorite. Nature. 1970; 228: p.923-926.

8. Kvenvolden KA, Lawless JG, Ponnamperuma C. Nonprotein Amino Acids in the Murchison Meteorite. Proceedings of the National Academy of Sciences. 1971; 68-2: p.486-490.

9. Cronin JR, Moore CB. Amino Acid Analyses of the Murchison, Murray and Allende Carbonaceous Chondrites. Science. 1971; 172: p.1327-1319.

10. Oró J, Gibert J, Lichtenstein H, Wikstrom S, Flory DA. Amino-acids, Aliphatic and Aromatic Hydrocarbons in the Murchison Meteorite. Nature. 1971; 230: p.105-106.

11. Engel MH, Nagy B. Distribution and enantiomeric composition of amino acids in the Murchison meteorite. Nature 1982; 296: p.837-840.

12. Furukawa Y, Sekine T, Oba M, Kakegawa T, Nakazawa H. Biomolecule formation by oceanic impacts on early Earth. Nature geoscience. 2009; 2: p. 62-66.

13. Takeuchi Y, Furukawa Y, Kobayashi T, Sekine T, Terada N, Kakegawa T. Impact-induced amino acid formation on Hadean Earth and Noachian Mars. Scientific Reports. 2020; 20: 9229.

14. Ajzenberg-Selove F. Energy Levels of Light Nuclei A = 13-15. Nuclear Physics A. 1991; 523: p. 1-196.

15. Poluianov SV, Kovaltsov GA, Mishev AL, Usoskin IG. Production of cosmogenic isotopes $^7$Be,

$^{10}$Be, $^{14}$C, $^{22}$Na, and $^{36}$Cl in the atmosphere: Altitudinal profiles of yield functions. Journal of Geophysical Research: Atmospheres. 2016; 121(13): p. 8125-8136.

16. Kanu AM, Comfort LL, Guilderson TP, Cameron-Smith PJ, Bergmann DJ, Atlas EL, Schuffler S, Boering KA. Measurements and modeling of contemporary radiocarbon in the stratosphere. Geophysical Research Letters. 2016; 43(3): p. 1399-1406.

17. Wilkinson DH. Evaluation of beta-decay Part I. The traditional phase space factors. Nuclear Instruments and Methods in Physics Research A. 1989; 275: p. 378-386.

18. Wilkinson DH. Evaluation of beta-decay Part II. Finite mass and size effects. Nuclear Instruments and Methods in Physics Research A. 1990; 290: p. 509-515.

19. Wilkinson DH. Evaluation of beta-decay Part III. The complex gamma function. Nuclear Instruments and Methods in Physics Research A. 1993; 335: p. 305-309.

20. Allen JS, Burman WB, Herrmannsfeldt WB, Stänhelin, Braid TH. Determination of the Beta-Decay Interaction from Electron-Neutrino Angular Correlation Measurements. Physical Review. 1959; 116: p. 134-143.

21. Johnson CH, Pleasonton F, Carlson TA. Precision Measurement of the Recoil Energy Spectrum from the Decay of He$^6$. Physical Review. 1963; 132: p. 1149-1165.

22. Frisch MJ, Trucks GW, Schlegel HB et al., Gaussian 16, Revision C.01, Gaussian, Inc., Wallingford CT, 2019.

23. Blanksby SJ, Ellison GB. Bond Dissociation Energies of Organic Molecules. Accounts of Chemical Research. 2003; 36: p. 255–263.

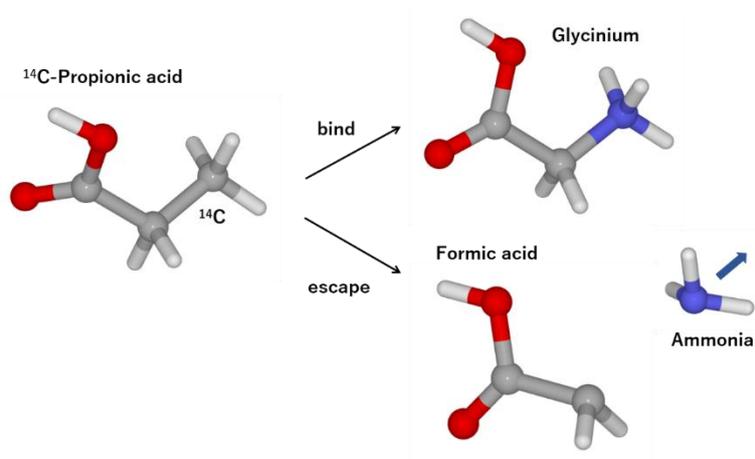

**Fig 1.** Schematic illustration of structure of a $^{14}$C-propionic acid (left) and glycinium production and nonproduction in the β-decay of $^{14}$C-propionic acid (right).

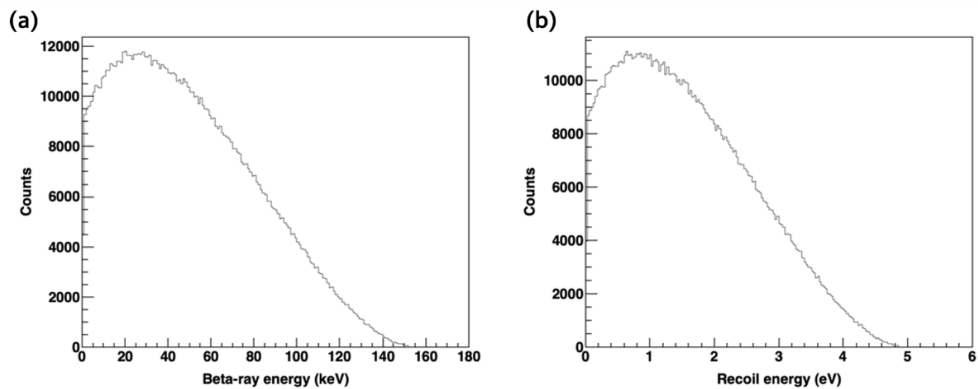

**Fig 2.** β-ray energy (a) and NH$_3$ recoil energy (b) distributions by a developed Monte simulation code with 1,000,000 event-by-event calculation.

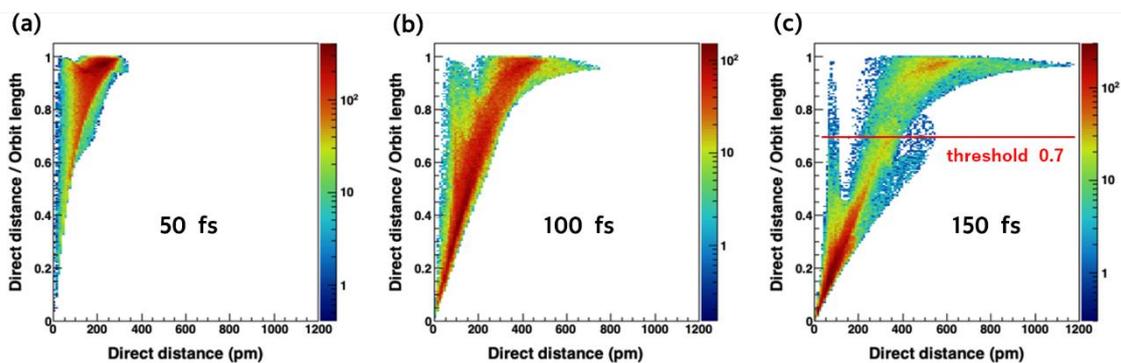

**Fig 3.** Two-dimensional plot for direct distance vs curving ratio of recoiled NH$_3$ calculated using developed Monte Carlo simulations with different elapsed times: (a) 50, (b) 100, and (c) 150 fs.

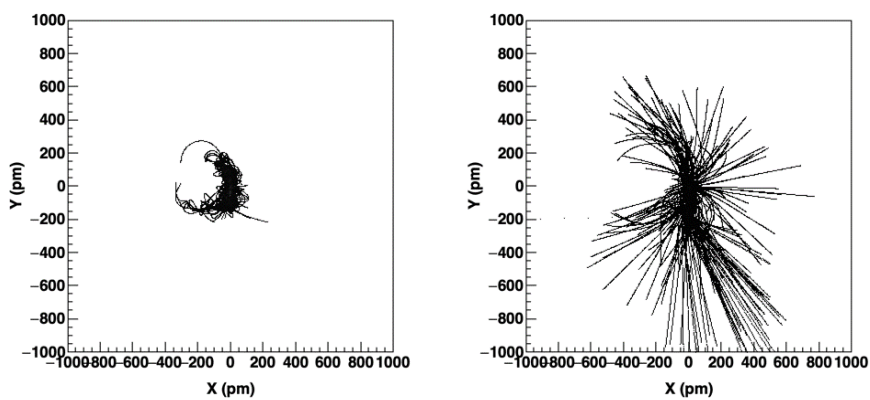

**Fig 4.** NH$_3$ orbits in xy-projections for first 100 events for event group (a) below and (b) above 0.7 of curving ratios.

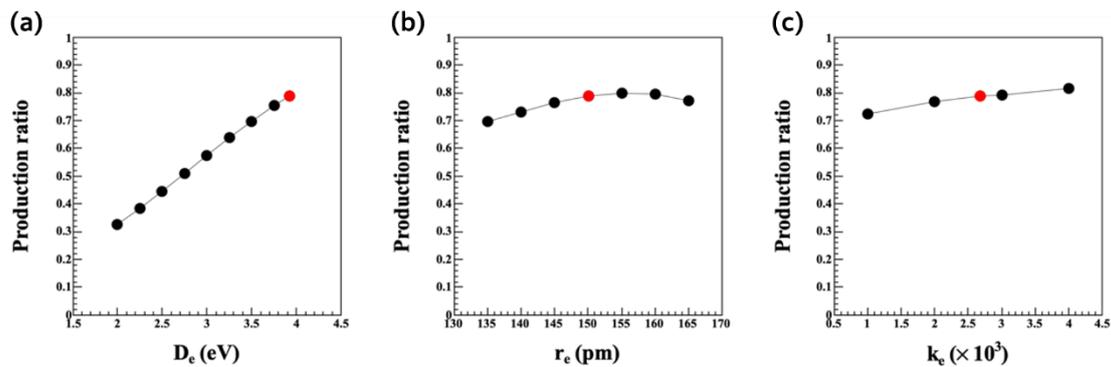

**Fig 5.** NH$_3$ remaining ratio, which is production ratio of glycinium from [3-$^{14}$C]propionic acid, in depth function of C–N binding potentials: (a) $D_e$, (b) $r_e$, and (c) $k_e$.

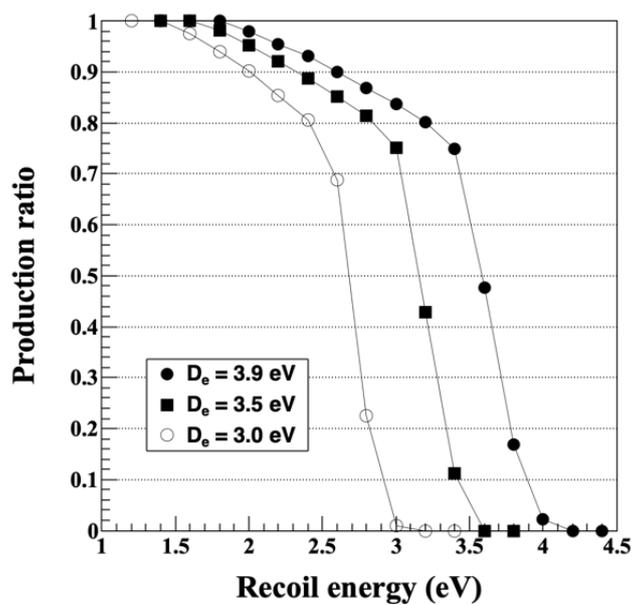

**Fig 6.** Glycinium production ratio as function of NH$_3$ recoil energy with potential depths of 3.9, 3.5, and 3.0 eV.